\documentclass[conference]{IEEEtran}
\usepackage{defpaper}
\usepackage{epsfig}
\usepackage{latexsym}

\newcommand{\filefig}[4]{
  \begin{figure}[htb]
  \vskip 1pt
    \begin{center}
      \setlength{\epsfxsize}{#4}
      \leavevmode
      \epsfbox {#2}
      \caption{\protect {#1}}
      \label {#3}
   \end{center}
\end{figure}}

      {
      \begin{minipage}
      {0.462\textwidth}\medskip\noindent
       \scriptsize\raggedright\sf\noindent
       {\bf Input:} #2\\
       {\bf Output:} #3\\
       \rule[1pt]{1.0\textwidth}{0.3pt}%
       \setlength{\leftmargini}{0em}%
       \setlength{\leftmarginii}{1.5em}%
       \setlength{\leftmarginiii}{1.5em}%
       \setlength{\leftmarginiv}{1.5em}%
       \setlength{\leftmarginv}{1.5em}%
       \begin{description}}
      {\end{description}\medskip
       \end{minipage}
       }

\long\def\comment#1{}

\long\def\symbolfootnote[#1]#2{\begingroup%
\def\thefootnote{\fnsymbol{footnote}}\footnote[#1]{#2}\endgroup}

\ifCLASSINFOpdf
\else
\fi
\begin{document}
%
\title{Managing Clouds in Cloud Platforms}

\author{\IEEEauthorblockN{Hassan Gobjuka}
\IEEEauthorblockA{Verizon\\
919 Hidden Ridge\\
Irving, TX 75083\\
Email: hasan.gobjuka@verizon.com}
}

\author{\begin{tabular}[t]{c@{\extracolsep{1em}}c@{\extracolsep{1em}}c}
           Kamal A. Ahmat & Hassan Gobjuka\\
\it Department of Information Technology & \it Verizon \\
\it City University of New York & \it 919 Hidden Ridge \\
\it New York, NY 11101 & \it Irving, TX 75038 \\
\it {\small\tt kamal.ahmat@live.lagcc.cuny.edu} & \it {\small\tt hasan.gobjuka@verizon.com} 
\end{tabular}}


%


\maketitle



%
\IEEEpeerreviewmaketitle

\section{Motivation}

Managing cloud services is a fundamental challenge in today's virtualized environments. These challenges equally face both providers and consumers of cloud services. The issue becomes even more challenging in virtualized environments that support mobile clouds. Cloud computing platforms such as Amazon EC2 provide customers with flexible, on demand resources at low cost. However, they fail to provide seamless infrastructure management and monitoring capabilities that many customers may need. For instance, Amazon EC2 doesn't fully support cloud services automated discovery and it requires a private set of authentication credentials. Salesforce.com, on the other hand, do not provide monitoring access to their underlying systems. Moreover, these systems fail to provide infrastructure monitoring of heterogenous and legacy systems that don't support agents.
In this work, we explore how to build a cloud management system that combines heterogeneous management of virtual resources with comprehensive management of physical devices. We propose an initial prototype for automated cloud management and monitoring framework. Our ultimate goal is to develop a framework that have the capability of automatically tracking configuration and relationships while providing full event management, measuring performance and testing thresholds, and measuring availability consistently. Armed with such a framework, operators can make better decisions quickly and more efficiently.

\comment{while existing offerings are useful for providing basic
computation and storage resources, they fail to provide the security
and network controls that many customers would like, you've probably found that legacy management tools fall short.
In the cloud, static really doesn't apply. A workload might be on one server one minute, on many the next, or on another in the next instant. This may or may not be apparent to the "developer" or service designer, or even the operator of the cloud, in fact its most likely not.
But how does one assemble a bunch of virtualized resources that need to be brought together to meet the needs of a service? Are all the sub-elements assembled via a model and "put on the stack" or do you do this at runtime? How long does this process take? How does a service know when to return its composite resources back to the "pool?" Does it have a default time to live? The model map of the resources need to ensure that they are providing the level of service required by the service.

}

\section{Challenges}
\label{model}

These tasks are achieved through an agentless monitoring of the cloud's infrastructure. While traditional network management methods suffer from inherited difficulties \cite{Benson,hbton}, implementing seamless network management and monitoring framework entails several new challenges:

\begin{itemize}
  \item Discovering the relationship of virtualized resources to underlying physical infrastructure.
  \item Minimizing the overhead of monitoring and problem determination across a physical and virtualized infrastructure.
  \item Handling security-related constraints that may affect data collection is probably one of the most serious issues. 
  \item Response action should be taken regarding a particular virtual or physical device within the hard response deadline time frame. In agentless-based monitoring systems, this can be insured only by implementing high number of threads, which in turn increases complexity.
  \item Dealing with infrastructure management issues such as root-cause analysis becomes more complex.
\end{itemize}

\comment{
\subsection{Challenges to Cloud service consumers}
\begin{itemize}
  \item \textbf{Limited visibility of management information from cloud resources} Many customers have also used Zenoss to help manage cloud-based services, where agent deployment is simply impossible. If a legacy system requires agents to function, and the cloud vendor won't allow that access, then you're stuck. Either. you're delivering services to end users and can't monitor availability and performance, or you install separate point products and attempt to integrate the solutions. Without the agent requirement, Zenoss allows you to simply include cloud services along with your network devices, real and virtualized servers.
  \item \textbf{Lack of management information from outsourced applications} In today's data center, traditional technical performance metrics may be unavailable or inappropriate. Cloud-based applications such as Salesforce.com generally do not provide monitoring access to their underlying systems. Customer-facing Web applications may deliver unacceptable performance even while underlying systems appear to be working successfully. What's needed is the ability to execute and measure the results of real transactions. Zenoss includes the ability to record and playback complex Web interactions and use the results for monitoring. You can monitor availability and performance of cloud-based applications, alert when response time thresholds are exceeded, and easily compare metrics from these synthetic transactions with metrics from your production systems. In addition to Web transactions, Zenoss can also test relational database queries and mail delivery.
  \item \textbf{Limited ability to install and use traditional management agents} is used to interact with the database. All changes made to the network layout by the user are reflected to the persistent tier.

  \item \textbf{Requirement to install additional management tools to achieve visibility} is used to interact with the database. All changes made to the network layout by the user are reflected to the persistent tier.
      
  \item \textbf{Dynamic expansion and contraction of resources taxes manual deployment and configuration processes} is used to interact with the database. All changes made to the network layout by the user are reflected to the persistent tier.
      
\end{itemize}

\subsection{Challenges to Cloud service providers}

\begin{itemize}
  \item \textbf{Agent-based management is impossible} Many customers have also used Zenoss to help manage cloud-based services, where agent deployment is simply impossible. If a legacy system requires agents to function, and the cloud vendor won't allow that access, then you're stuck. Either. you're delivering services to end users and can't monitor availability and performance, or you install separate point products and attempt to integrate the solutions. Without the agent requirement, Zenoss allows you to simply include cloud services along with your network devices, real and virtualized servers.
  \item \textbf{Synthetic Transactions} In today's data center, traditional technical performance metrics may be unavailable or inappropriate. Cloud-based applications such as Salesforce.com generally do not provide monitoring access to their underlying systems. Customer-facing Web applications may deliver unacceptable performance even while underlying systems appear to be working successfully. What's needed is the ability to execute and measure the results of real transactions. Zenoss includes the ability to record and playback complex Web interactions and use the results for monitoring. You can monitor availability and performance of cloud-based applications, alert when response time thresholds are exceeded, and easily compare metrics from these synthetic transactions with metrics from your production systems.
In addition to Web transactions, Zenoss can also test relational database queries and mail delivery.
  \item \textbf{Persistent Tier} is used to interact with the database. All changes made to the network layout by the user are reflected to the persistent tier.
\end{itemize}
}
\section{Design and Implementation}

We propose an event-based model where events are placed on an in-memory publish/subscribe bus on the Management Server, enabling a high throughput of events.

The event bus architecture, depicted in Figure \ref{arch} enables any authorized "mediator" to create events on the bus, and any authorized "consumer" to access events from the bus. Events on the bus show current status of infrastructure components.

\filefig{An initial prototype of our cloud management and monitoring system.}
{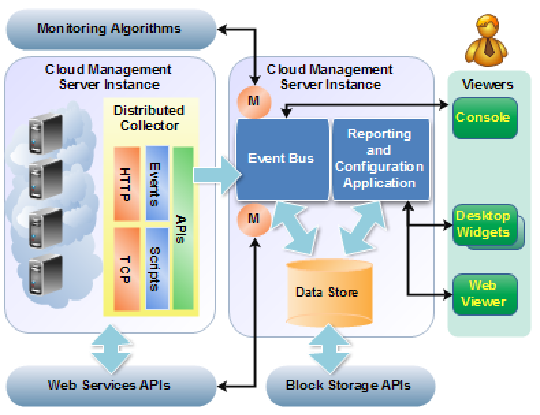}    
{arch}      
{3.1in}        

The framework will provide a set of APIs to simplify creation of consumer and mediator applications. A set of language extensions and Web services will be used to enable Perl, Ruby, or Java scripts to create events on the bus. To support high level of reliability and scalability, the Distributed Collector subsystem will be multi-threaded. Furthermore, events will are normalized from any source into a common format, which will enable consistent processing.


\end{document}